\documentstyle[12pt,graphicx,caption2]{article}

\begin{document}

\title{$n\bar{n}$ transitions in medium}
\author{V.I.Nazaruk\\
Institute for Nuclear Research of RAS, 60th October\\
Anniversary Prospect 7a, 117312 Moscow, Russia.*}

\date{}
\maketitle
\bigskip

\begin{abstract}
The models of $n\bar{n}$ transitions in the medium based on unitary $S$-matrix are 
considered. The time-dependence and corrections to the models are studied. The
lower limits on the free-space $n\bar{n}$ oscillation time are obtained as well. 
\end{abstract}

\vspace{5mm}

\vspace{5mm}

\vspace{1cm}

*E-mail: nazaruk@inr.ru

\newpage
\setcounter{equation}{0}
\section{Introduction}
In the standard calculations of $ab$ oscillations in the medium [1-3] the interaction of 
particles $a$ and $b$ with the matter is described by the potentials $U_{a,b}$ (potential 
model). ${\rm Im}U_b$ is responsible for loss of $b$-particle intensity. In particular, 
this model is used for the $n\bar{n}$ transitions in a medium [4-10] followed by annihilation:
\begin{equation}
n\rightarrow \bar{n}\rightarrow M,
\end{equation}
here $M$ are the annihilation mesons.

In [9] it was shown that one-particle model mentioned above does not describe the
total $ab$ (neutron-antineutron) transition probability as well as the channel corresponding 
to absorption of the $b$-particle (antineutron). The effect of final state absorption
(annihilation) acts in the opposite (wrong) direction, which tends to the additional
suppression of the $n\bar{n}$ transition. The $S$-matrix should be unitary.

In [11] we have proposed the model based on the diagram technique which does not contain
the non-hermitian operators. Subsequently, this calculation was repeated in [12]. However, in 
[13] it was shown that this model is unsuitable: the neutron line entering into the $n\bar{n}$ 
transition vertex should be the wave function, but not the propagator, as in the model based 
on the diagram technique. For the problem under study this fact is crucial. It leads 
to the cardinal error for the process in nuclei. The $n\bar{n}$ transitions in the medium and 
vacuum are not reproduced at all. If the neutron binding energy goes to zero, the result
diverges (see Eqs. (18) and (19) of Ref. [11] or Eqs. (15) and (17) of Ref. [12]). So we 
abandoned this model [13]. (In the recent manuscript [14] the previous calculations [11,12] 
have been repeated. The model and calculation are the same as in [11,12]. Unfortunately,
several statements are erroneous [15], in particular, the conclusion based on an analogy 
with the nucleus form-factor at zero momentum transfer (for more details, see [15]).)

In [16] the model which is free of drawbacks given above has been proposed (model {\bf b}
in the notations of present paper). However, the consideration was schematic since our 
concern was only with the role of the final state absorption in principle. In Sect. 2 this 
model as well as the model with bare propagator are studied in detail. The corrections to 
the models (Sect. 3) and time-dependence (Sect. 4) are considered as well. In addition, we 
sum up the present state of the investigations of this problem (Sect. 5).

The basic material is given in Sects. 2 and 5.

\section{Models}
First of all we consider the antineutron annihilation in the medium. The annihilation 
amplitude $M_a$ is defined as 
\begin{equation}
<\!f0\!\mid T\exp (-i\int dx{\cal H}(x))-1\mid\!0\bar{n}_{p}\!>=
N(2\pi )^4\delta ^4(p_f-p_i)M_a.
\end{equation}
Here ${\cal H}$ is the Hamiltonian of the $\bar{n}$-medium interaction, $\mid\!0\bar{n}_{p}\!>$ 
is the state of the medium containing the $\bar{n}$ with the 4-momentum $p=(\epsilon 
,{\bf p})$; $<\!f\!\mid $ denotes the annihilation products, $N$ includes the normalization 
factors of the wave functions. The antineutron annihilation width $\Gamma $ is expressed 
through $M_a$:
\begin{equation}
\Gamma \sim \int d\Phi \mid\!M_a\!\mid ^2.
\end{equation}

For the Hamiltonian ${\cal H}$ we consider the model
\begin{eqnarray}
{\cal H}={\cal H}_a+V\bar{\Psi }_{\bar{n}}\Psi _{\bar{n}},\nonumber\\
H(t)=\int d^3x{\cal H}(x)=H_a(t)+V,
\end{eqnarray}
where ${\cal H}_a$ is the effective annihilation Hamiltonian in the second quantization
representation, $V$ is the residual scalar field. The diagrams for the model (4) are shown 
in Fig. 1. The first diagram corresponds to the first order in ${\cal H}_a$ and so on.

\begin{figure}[h]
  {\includegraphics[height=.25\textheight]{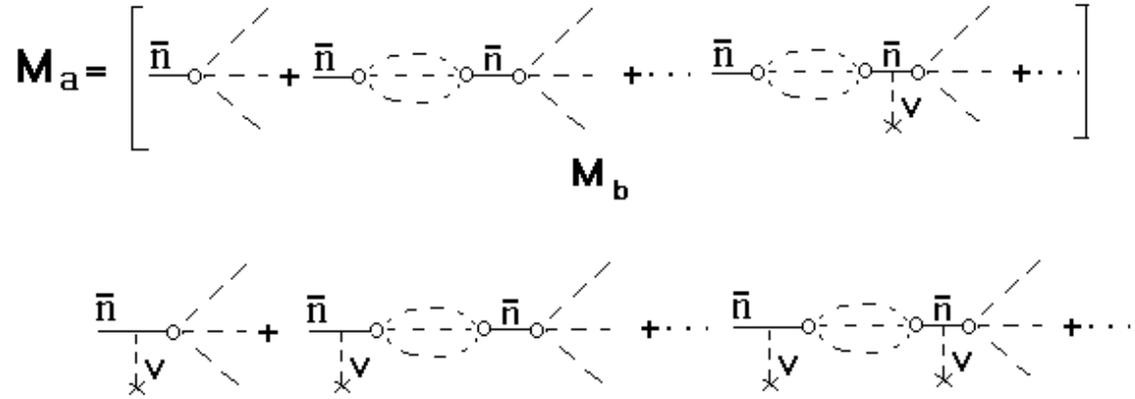}}
  \caption{Antineutron annihilation in the medium. The annihilation is shown by a circle}
\end{figure}

Consider now the process (1). The neutron wave function is 
\begin{equation}
n(x)=\Omega ^{-1/2}\exp (-ipx).
\end{equation}
Here $p=(\epsilon ,{\bf p})$ is the neutron 4-momentum; $\epsilon ={\bf p}^2/2m+U_n$,
where $U_n$ is the neutron potential. The interaction Hamiltonian has the form
\begin{equation}
H_I=H_{n\bar{n}}+H,
\end{equation}
\begin{equation}
H_{n\bar{n}}(t)=\int d^3x(\epsilon _{n\bar{n}}\bar{\Psi }_{\bar{n}}(x)\Psi _n(x)+H.c.)
\end{equation}
Here $H_{n\bar{n}}$ is the Hamiltonian of $n\bar{n}$ conversion [6], $\epsilon _{n\bar{n}}$ 
is a small parameter with $\epsilon _{n\bar{n}}=1/\tau $, where $\tau $ is the free-space 
$n\bar{n}$ oscillation time; $m_n=m_{\bar{n}}=m$. In the lowest order in $H_{n\bar{n}}$ the 
amplitude of process (1) is {\em uniquely} determined by the Hamiltonian (6):
\begin{equation}
M=\epsilon _{n\bar{n}}G_0M_a,
\end{equation}
\begin{equation}
G_0=\frac{1}{\epsilon -{\bf p}^2/2m-U_n+i0},
\end{equation}
${\bf p}_{\bar{n}}={\bf p}$, $\epsilon _{\bar{n}}=\epsilon $. Here $G_0$ is the antineutron 
propagator. The corresponding diagram is shown in Fig. 2a. The annihilation amplitude $M_a$ 
is given by (2), where ${\cal H}={\cal H}_a+V\bar{\Psi }_{\bar{n}}\Psi _{\bar{n}}$. Since 
$M_a$ contains all the $\bar{n}$-medium interactions followed by annihilation including 
antineutron rescattering in the initial state, the antineutron propagator $G_0$ is bare. Once 
the antineutron annihilation amplitude is defined by (2), the expression for the process 
amplitude (8) {\em rigorously follows} from (6). For the time being we do not go into the 
singularity $G_0\sim 1/0$.

\begin{figure}[h]
  {\includegraphics[height=.25\textheight]{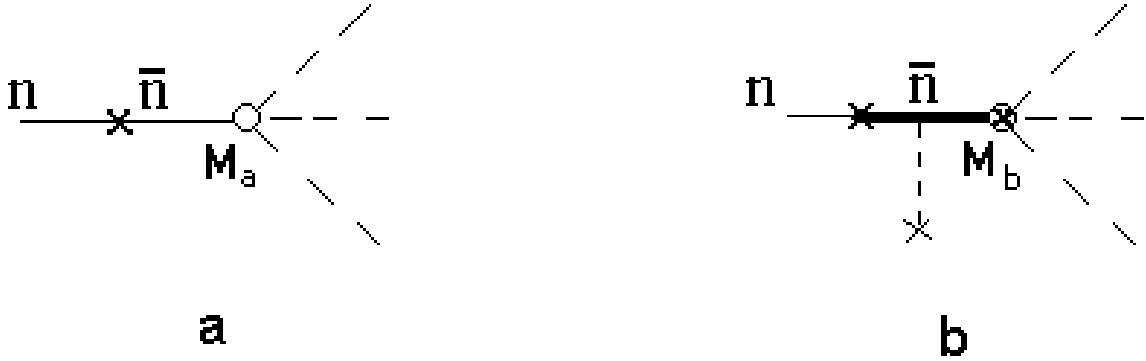}}
  \caption{{\bf a} $n\bar{n}$ transition in the medium followed by annihilation. The 
antineutron annihilation is shown by a circle. {\bf b} Same as {\bf a} but the antineutron
propagator is dressed (see text)}
\end{figure}

One can construct the model with the dressed propagator. We include the scalar field $V$ 
in the antineutron Green function
\begin{equation}
G_d=G_0+G_0VG_0+...=\frac{1}{(1/G_0)-V}=-\frac{1}{V}=-\frac{1}{\Sigma },
\end{equation}
$\Sigma =V$, where $\Sigma $ is the antineutron self-energy. The process amplitude is
\begin{equation}
M=\epsilon _{n\bar{n}}G_dM_b,
\end{equation}
$G_dM_b=G_0M_a$ (see Fig. 2b). The block in the square braces shown in Fig.1 corresponds 
to the vertex function $M_b$. The models shown in Figs. 2a and 2b we denote as the models 
{\bf a} and {\bf b}, respectively.

In both models the interaction Hamiltonians $H_I$ and unperturbed Hamiltonians are the same.
If $\Sigma \rightarrow 0$, the model {\bf b} goes into model {\bf a}. In this sense the model 
{\bf a} is the limiting case of the model {\bf b}.

We consider the model {\bf b}. For the process width $\Gamma _b$ one obtains
\begin{equation}
\Gamma _b=N_1\int d\Phi \mid\!M\!\mid ^2=\frac{\epsilon _{n\bar{n}}^2}{\Sigma ^2}N_1\int d\Phi
\mid\!M_b\!\mid ^2=\frac{\epsilon _{n\bar{n}}^2}{\Sigma ^2}\Gamma ',
\end{equation}
\begin{equation}
\Gamma '=N_1\int d\Phi \mid\!M_b\!\mid ^2,
\end{equation}
where  $\Gamma '$ is the annihilation width of $\bar{n}$ calculated through the $M_b$ (and 
not $M_a$). The normalization multiplier $N_1$ is the same for $\Gamma _b$ and $\Gamma '$. 
The vertex function $M_b$ is unknown. (We recall the antineutron annihilation width 
$\Gamma $ is expressed through the amplitude $M_a$.) For the estimation we put
\begin{equation}
M_b=M_a, \quad \Gamma '=\Gamma.
\end{equation}
This is an uncontrollable approximation.

The time-dependence is determined by the exponential decay law:
\begin{equation}
W_b(t)=1-e^{-\Gamma _bt}\approx \Gamma _bt=\frac{\epsilon _{n\bar{n}}^2}{\Sigma ^2}\Gamma t.
\end{equation}
Equation (15) illustrates the result sensitivity to the value of parameter $\Sigma $.

On the other hand, for $n\bar{n}$ transitions in nuclear matter the standard calculation
gives the inverse $\Gamma $-dependence [6-9] 
\begin{equation}
W_{{\rm stan}}(t)=2\epsilon _{n\bar{n}}^2t\frac{\Gamma /2}{({\rm Re}U_{\bar{n}}-U_n)^2+
(\Gamma /2)^2}\approx \frac{4\epsilon _{n\bar{n}}^2t}{\Gamma },
\end{equation}
where $U_{\bar{n}}$ is the antineutron optical potential. The wrong $\Gamma $-dependence is 
a direct consequence of the inapplicability of the model based on optical potential for the 
calculation of the total process probability [9]. (The above-mentioned model describes the 
probability of finding an antineutron only.)
 
Comparing with (15), one obtains
\begin{equation}
r= \frac{W_b}{W_{{\rm stan}}}=\frac{\Gamma ^2}{4\Sigma ^2}=25,
\end{equation}
where the values $\Gamma =100$ MeV and $\Sigma ={\rm Re}U_{\bar{n}}-U_n=10$ MeV have been 
used. Strictly speaking, the parameter $\Sigma $ is uncertain. We have put $\Sigma =
{\rm Re}U_{\bar{n}}-U_n=10$ MeV only for estimation.

The model {\bf b} leads to an increase of the $n\bar{n}$ transition probability. The lower 
limit on the free-space $n\bar{n}$ oscillation time $\tau ^b_{{\rm min}}$ increases as well:
\begin{equation}
\tau ^b_{{\rm min}}=(3.5-7.5)\cdot 10^{8}\; {\rm s}.
\end{equation}
This limit exceeds the previous one (see, for example, Refs. [5-8]) by a factor of five.
If $\Sigma \rightarrow 0$, $W_b$ rises quadratically. So Eq. (18) can be considered as
the estimation from below.

We return to the model shown in Fig. 2a. We use the basis $(n,\bar{n})$. The results do not 
depend on the basis. A main part of existing calculations have been done in $n-\bar{n}$ 
representation. The physics of the problem is in the Hamiltonian. The transition to the basis 
of stationary states is a formal step. It has a sense only in the case of the potential model 
$H=H_{{\rm pot}}={\rm Re}U_{\bar{n}}-U_n-i\Gamma /2=$const., when the Hamiltonian of 
$\bar{n}$-medium interaction is replaced by the effective mass $H\rightarrow H_{{\rm pot}}=
m_{{\rm eff}}$ because the Hermitian Hamiltonian of interaction of the stationary states with 
the medium is unknown. Since we work beyond the potential model, the procedure of 
diagonalization of mass matrix is unrelated to our problem.

The amplitude (8) diverges
\begin{equation}
M=\epsilon _{n\bar{n}}G_0M_a\sim \frac{1}{0}.
\end{equation}
(See also Eq. (21) of Ref. [13].) These are infrared singularities conditioned by zero 
momentum transfer in the $n\bar{n}$ transition vertex. (In the model {\bf b} the effective 
momentum transfer $q_0=V=\Sigma $ takes place.) 

For solving the problem the field-theoretical approach with finite time 
interval [17] is used. It is infrared free. If $H=H_{{\rm pot}}$, the approach with finite 
time interval reproduces all the results on the particle oscillations, in particular, the 
$n\bar{n}$ transition with $\bar{n}$ in the final state. (Recall that our purpose is to 
describe the process (1) by means of Hermitian Hamiltonian.)

For the model {\bf a} the process (1) probability was found to be [10,13]
\begin{equation}
W_a(t)\approx  W_f(t)=\epsilon _{n\bar{n}}^2t^2, \quad \Gamma t\gg 1,
\end{equation}
where $W_f$ is the free-space $n\bar{n}$ transition probability. Owing to annihilation 
channel, $W_a$ is practically equal to the free-space $n\bar{n}$ transition probability. 
If $t\rightarrow \infty $, Eq. (20) diverges just as the modulus (19) squared does. If 
$\Sigma \rightarrow 0$, Eq. (15) diverges quadratically as well.

The explanation of the $t^2$-dependence is simple. The process shown in Fig. 2a represents 
two consecutive subprocesses. The speed and probability of the whole process are defined 
by those of the slower subprocess. If $1/\Gamma \ll t$, the annihilation can be considered 
instantaneous. So, the probability of process (1) is defined by the speed of the $n\bar{n}$ 
transition: $W_a\approx W_f\sim t^2$. 

Distribution (20) leads to very strong restriction on the free-space $n\bar{n}$ oscillation 
time [10,13]:
\begin{equation}
\tau ^a_{{\rm min}}=10^{16}\; {\rm yr}.
\end{equation}

\section{Corrections}
We show that for the $n\bar{n}$ transition in medium the corrections to the models 
and additional baryon-number-violating processes (see Fig. 3) cannot essentially change the 
results. First of all we consider the incoherent contribution of the diagrams 3. In Fig. 3a 
a meson is radiated before the $n\bar{n}$ transition. The interaction Hamiltonian has the form
\begin{equation}
H_I=\int d^3xg\Psi ^+_n\Phi \Psi _n+H_{n\bar{n}}+H.
\end{equation}
In the following the background neutron potential is omitted. The neutron wave function is 
given by (5), were $p=(p_0,{\bf p})$ and $p_0=m+{\bf p}^2/2m$.

For the process amplitude $M_{3a}$ one obtains
\begin{equation}
M_{3a}=gG\epsilon _{n\bar{n}}GM^{(n-1)},
\end{equation}
\begin{equation}
G=\frac{1}{p_0-q_0-m-({\bf p}-{\bf q})^2/2m+i0},
\end{equation}
where $q$ is the 4-momentum of meson radiated, $M^{(n-1)}$ is the amplitude of antineutron
annihilation in the medium in the $(n-1)$ mesons. As with model {\bf a}, the antineutron 
propagator $G$ is {\em bare}; the $\bar{n}$ self-energy $\Sigma =0$. (The same is true
for Figs. 3b-3d.)

\begin{figure}[h]
  {\includegraphics[height=.3\textheight]{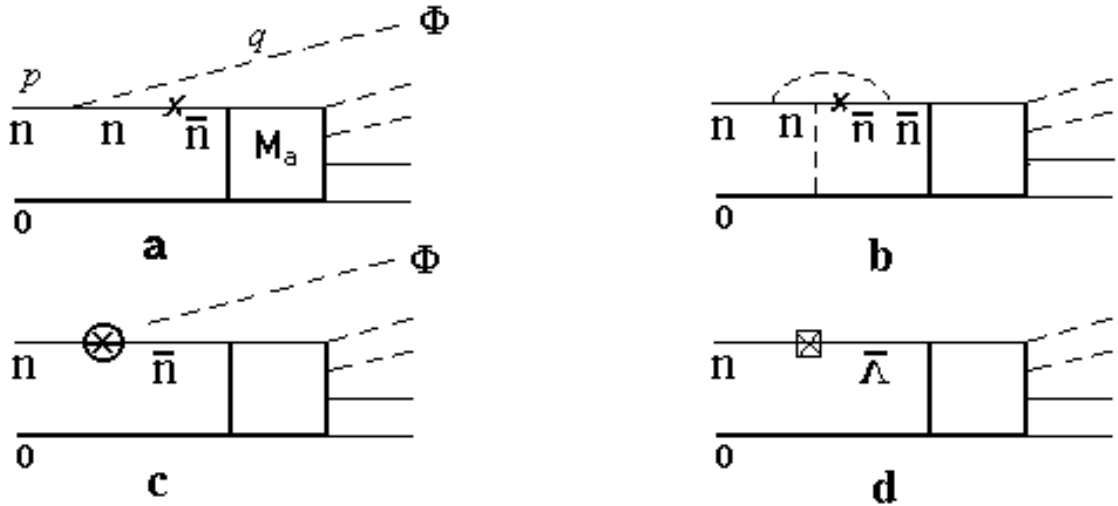}}
  \caption{Corrections to the models ({\bf a} and {\bf b}) and
additional baryon-number-violating processes ({\bf c} and {\bf d})}
\end{figure}

If $q\rightarrow 0$, the amplitude $M_{3a}$ increases since $G\rightarrow G_s$,
\begin{equation}
G_s=\frac{1}{p_0-m-{\bf p}^2/2m}\sim \frac{1}{0}.
\end{equation}
(The limiting transition $q\rightarrow 0$ for the diagram 3a is an imaginary procedure 
because in the vertex $n\rightarrow n\Phi $ the real meson is escaped and so $q_0\geq 
m_{\Phi }$.) The fact that the amplitude increases is essential for us because for Fig. 
2a $q=0$. Due to this $G_0\sim 1/0$ and $W_a\gg W_b$.

Let $\Gamma _{3a}$ and $\Gamma ^{(n)}$ be the widths corresponding to the Fig. 3a
and annihilation width of $\bar{n}$ in the $(n)$ mesons, respectively; $\Gamma =
\sum_{(n)}\Gamma ^{(n)}$. Taking into account that $\Gamma ^{(n)}$ is a smooth
function of $\sqrt{s}$ and summing over $(n)$, it is easy to get the estimation:
\begin{equation}
\Gamma _{3a}\approx 5\cdot 10^{-3}g^2\frac{\epsilon _{n\bar{n}}^2}{m^2_{\Phi }}\Gamma
\approx \frac{\epsilon _{n\bar{n}}^2}{m^2_{\Phi }}\Gamma .
\end{equation}

The time-dependence is determined by the exponential decay law:
\begin{equation}
W_{3a}(t)\approx \Gamma _{3a}t=\frac{\epsilon _{n\bar{n}}^2}{m^2_{\Phi }}\Gamma t.
\end{equation}
Comparing with (15) we have: $W_{3a}/W_b=V^2/m^2_{\Phi }\ll 1$. So for the model {\bf b}
the contribution of diagram 3a is negligible.

For the model {\bf a} the contribution of diagram 3a is inessential as well. Indeed,
using Eqs. (27) and (20) we get
\begin{equation}
\frac{W_{3a}(t)}{W_a(t)}=\frac{\Gamma }{m^2_{\pi }t},
\end{equation}
where we have put $m_{\Phi }=m_{\pi }$. Consequently, if
\begin{equation}
m^2_{\pi }t/\Gamma \gg 1,
\end{equation}
and
\begin{equation}
\Gamma t\gg 1
\end{equation}
(see (20)) then the contribution of diagram 3a is negligible. For the $n\bar{n}$ 
transition in nuclei these conditions are fulfilled since in this case $\Gamma \sim 100$ 
MeV and $t=T_0=1.3$ yr, where $T_0$ is the observation time in proton-decay type experiment 
[18].) In fact, it is suffice to hold condition (30) only because it is more strong.

In the calculations made above the free-space $n\bar{n}$ transition operator has
been used. This is impulse approximation which is employed for nuclear $\beta $ decay,
for instance. The simplest medium correction to the vertex (or off-diagonal mass, or
transition mass) is shown in Fig. 3b. In this event the replacement should be made: 
\begin{equation}
\epsilon _{n\bar{n}}\rightarrow \epsilon _m=\epsilon _{n\bar{n}}(1+\Delta \epsilon ),
\end{equation}
$\Delta \epsilon  =\epsilon _{3b}/\epsilon _{n\bar{n}}$, where $\epsilon _{3b}$ is the 
correction to $\epsilon _{n\bar{n}}$ produced by the diagram 3b. For the model {\bf a} the 
limit becomes
\begin{equation}
\tau ^a_{{\rm min}}=(1+\Delta \epsilon )10^{16}\; {\rm yr}.
\end{equation}
Obviously, the $\Delta \epsilon $ cannot change the order of magnitude of $\tau 
_{{\rm min}}$ since the $n\rightarrow \bar{n}$ operator is essentially zero-range one. 
The free-space $n\bar{n}$ transition comes from the exchange of Higgs bosons with the 
mass $m_H>10^5$ GeV [5]. Since $m_H\gg m_W$ ($m_W$ is the mass of $W$-boson), the
renormalization effects should not exceed those characteristic of nuclear $\beta $
decay which is less than 0.25 [19]. So the medium corrections to the vertex are 
inessential for us. The same is true for the model {\bf b}.

Consider now the baryon-number-violating decay $n\rightarrow \bar{n}\Phi $ [20]
shown in Fig. 3c. It leads to the same final state, as the processes depicted in
Figs. 2, 3a and 3b. Denoting $\mid {\bf q}\mid=q$, for the decay width $\Gamma _{3c}$ 
one obtains
\begin{equation}
\Gamma _{3c}=\frac{\epsilon _{\Phi }^2}{(2\pi )^2}\int dq\frac{q^2}{q_0}G^2\Gamma (q),
\end{equation}
$q_0^2=q^2+m_{\Phi }^2$. The parameter $\epsilon _{\Phi }$ corresponding to the vertex
$n\rightarrow \bar{n}\Phi $ is unknown and so no detailed calculation is possible.

The baryon-number-violating conversion $n\rightarrow \bar{\Lambda }$ in the medium [20] 
shown in Fig. 3d cannot produce interference, since it contains $K$-meson in the final 
state. For the rest of the diagrams the significant interferences are unlikely because 
the final states in $\bar{n}N$ annihilation are very complicated configurations and 
persistent phase relations between different amplitudes cannot be expected. This 
qualitative picture is confirmed by our calculations [21] for $\bar{p}$-nuclear annihilation.
It is easy to verify the following statement: if the incoherent contribution of the 
diagrams 3a-3c to the total nuclear annihilation width is taken into account, the
lower limit on the free-space $n\bar{n}$ oscillation time $\tau _{{\rm min}}$
becomes even better.

To summarise, the contribution of diagrams 3 is inessential for us. 

\section{Time-dependence}
The non-trivial circumstance is the quadratic time-dependence in the model {\bf a}: $W_a\sim 
t^2$. The heart of the problem is as follows. The processes depicted by the diagrams 2b and 3 
are described by the exponential decay law. In the first vertex of these diagrams the
momentum transfer (Figs. 3a-3c), or effective momentum transfer (Figs. 2b, 3d) takes place.
The diagram 2a contains the infrared divergence conditioned by zero momentum transfer in the 
$n\bar{n}$ transition vertex. This is unremovable perculiarity. This means that the standard 
$S$-matrix approach is inapplicable [10,13,17]. In such an event, the other surprises can be 
expected as well. From this standpoint a non-exponential behaviour comes as no surprise to us. 
It seems natural that for non-singular and singular diagrams the functional structure of the 
results is different, including the time-dependence. The opposite situation would be strange.

The fact that for the processes with $q=0$ the $S$-matrix problem formulation $(\infty,
-\infty)$ is physically incorrect can be seen even from the limiting case $H=0$: if $H_I=
H_{n\bar{n}}$ (see (6)), the solution is periodic. It is obtained by means of non-stationary 
equations of motion and not $S$-matrix theory. To reproduce the limiting case $H\rightarrow 
0$, i.e. the periodic solution, we have to use the approach with finite time interval. 

If the problem is formulated on the interval $(t,0)$, the decay width $\Gamma $ cannot
be introduced since $\Gamma =\sum_{f\neq i}\mid S_{fi}(\infty,-\infty)\mid ^2/T_0$,
$T_0\rightarrow \infty $. This means that the standard calculation scheme should be
completely revised. (We would like to emphasise this fact.) The direct calculation by 
means of evolution operator gives the distribution (20).

Formally, the different time-dependence is due to $q$-dependence of amplitudes. We consider
Eq. (23), for instance. If $q$ decreases, the amplitude $M_{3a}$ increase; in the limit 
$q\rightarrow 0$ it is singular (see (25)). The point $q=0$ corresponds to realistic process 
shown in Fig. 2a. The $t^2$-dependence of this process is the consequence of the zero 
momentum transfer.

The more physical explanation of the $t^2$-dependence is as follows. In the Hamiltonian
(22) corresponding to Fig. 3a we put $H=H_{n\bar{n}}=0$. Then the virtual decay $n\rightarrow 
n\Phi $ takes place. The first vertex of the diagram 3a dictates the exponential decay law 
of the overall process shown in Fig. 3a. Similarly, in the Hamiltonian (6) corresponding to 
Fig. 2a, we put $H=0$. Then the free-space $n\bar{n}$ transition takes place which is 
quadratic in time: $W_f(t)=\epsilon _{n\bar{n}}^2t^2$. The first vertex determines the 
time-dependence of the whole process at least for small $\Gamma $. We also recall that even 
for proton decay the possibility of non-exponential behaviour is realistic [22-24].

\section{Discussian and summary}
The sole physical distinction between models {\bf a} and {\bf b} is the zero antineutron 
self-energy in the model {\bf a}; or, similarly, the definition of antineutron 
annihilation amplitude. However, it leads to the fundamentally different results.
 
If $\Sigma \rightarrow 0$, $W_b(t)$ diverges quadratically. This circumstance should be 
clarified; otherwise the model {\bf b} can be rejected. The calculation in the framework 
of the model {\bf a} gives the finite result, which justifies our approach from a conceptual 
point of view and consideration of the model {\bf a} at least as the limiting case. In reality 
the model {\bf a} seems quite realistic in itself. Indeed, we list the main drawbacks of the 
model {\bf b}.

1) The approximation $M_b=M_a$ is an uncontrollable one. The value of $\Sigma $ is uncertain.
These points are closely related.

2) The diagram 2b means that the annihilation is turned on upon forming of the self-energy 
part $\Sigma =V$ (after multiple rescattering of $\bar{n}$). This is counter-intuitive 
since at low energies [25,26]
\begin{equation}
\sigma _a>2.5\sigma _s,
\end{equation}
where $\sigma _a$ and $\sigma _s$ are the cross sections 
of free-space $\bar{n}N$ annihilation and $\bar{n}N$ scattering, respectively. The inverse 
picture is in order: in the first stage of the $\bar{n}$-medium interaction the annihilation 
occurs. This is obvious for the $n\bar{n}$ transitions in the gas. The model {\bf a} reproduces
the {\em competition} between scattering and annihilation in the intermediate state [27]. 

3) The time-dependence is a more important characteristic of any process. It is common 
knowledge that the $t$-dependence of the process probability in the vacuum and medium is 
identical (for example, exponential decay law (15)). In the model {\bf a} the $t$-dependencies 
in the vacuum and medium coincide: $W_a\sim t^2$ and $W_f\sim t^2$. The model {\bf b} gives 
$W_b\sim t$, whereas $W_f\sim t^2$. There is no reason known why we have such a fundamental 
change.

4) If $H=U_{\bar{n}}$, the model {\bf a} reproduces all the well-known results on particle 
oscillations [10] in contrast to the model {\bf b}.

The model {\bf a} is free of drawbacks given above. The physics of the model is absolutely 
standard. For instance, for the processes shown in Fig. 3 the antineutron propagators are 
bare as well. 

However, there is fundamental problem in the model {\bf a}: the singularity 
of the amplitude (19). The approach with finite time interval gives the finite result, 
which justifies the models {\bf a} and {\bf b} at least in principle.
Nevertheless, the time-dependence $W_a\sim t^2$ and limit (21) seem very unusual. The 
corresponding calculation contains too many new elements. Due to this we view the results 
of the model {\bf a} with certain caution. Besides, due to the zero momentum transfer in the 
$n\bar{n}$-transition vertex, the model is extremely sensitive to the $\Sigma $. The process 
under study is {\em unstable}. The small change of antineutron self-energy $\Sigma =0
\rightarrow \Sigma =V\neq 0$, or, similarly, effctive momentum transfer in the $n\bar{n}$ 
transition vertex converts the model {\bf a} to the model {\bf b}: $W_a\rightarrow W_b$ 
with $W_b\ll W_a$. (For the processes with non-zero momentum transfer the result is little 
affected by small change of $q$.) Although we don't see the specific reasons for similar 
scenario, it must not be ruled out. This is a point of great nicety.

Finally, the values $\tau ^b_{{\rm min}}=(3.5-7.5)\cdot 10^{8}$ s and $\tau ^a_{{\rm min}}=
10^{16}$ yr are interpreted as the estimations from below (conservative limit) and from above,
respectively. Further investigations are desirable.


\newpage
\begin{thebibliography}{99}
\bibitem{1}
M.L. Good, Phys. Rev. {\bf 106}, 591 (1957).
%
\bibitem{2}
M.L. Good, Phys. Rev. {\bf 110}, 550 (1958). 
%
\bibitem{3}
E. D. Commins and P. H. Bucksbaum, {\em Weak Interactions of Leptons and Quarks}
(Cambridge University Press, 1983).
%
\bibitem{4}
V. A. Kuzmin, JETF Lett. {\bf 12}, 228 (1970).  
%
\bibitem{5}
R. N. Mohapatra and R. E. Marshak, Phys. Rev. Lett. {\bf 44}, 1316 (1980).  
%
\bibitem{6}
K. G. Chetyrkin, M. V. Kazarnovsky, V. A. Kuzmin and M. E. Shaposhnikov,
Phys. Lett. B {\bf 99}, 358 (1981).
%
\bibitem{7}
J. Arafune and O. Miyamura, Prog. Theor. Phys. {\bf 66}, 661 (1981).
%
\bibitem{8}
W. M. Alberico {\it et al.}, Nucl. Phys. A {\bf 523}, 488 (1991).  
%
\bibitem{9}
V. I. Nazaruk, Eur. Phys. J. A {\bf 31}, 177 (2007).  
%
\bibitem{10}
V. I. Nazaruk, Eur. Phys. J. C {\bf 53}, 573 (2008).
%
\bibitem{11}
V. I. Nazaruk, Yad. Fiz. {\bf 56}, 153 (1993).
%
\bibitem{12}
L. A. Kondratyuk, Pis'ma Zh. Exsp. Theor. Fiz. {\bf 64}, 456 (1996).
%
\bibitem{13}
V. I. Nazaruk, Phys. Rev. C {\bf 58}, R1884 (1998)
%
\bibitem{14}
V. B. Kopeliovich, arXiv: 0912.5065.
%
\bibitem{15}
V. I. Nazaruk, arXiv: 1003.4360.
%
\bibitem{16}
V. I. Nazaruk, Mod. Phys. Lett. A {\bf 21}, 2189 (2006).
%
\bibitem{17}
V. I. Nazaruk, Phys. Lett. B {\bf 337}, 328 (1994).
%
\bibitem{18}
H. Takita {\it et al.}, Phys. Rev. D {\bf 34}, 902 (1986).
%
\bibitem{19}
B. Buch and S. M. Perez, Phys. Rev. Lett. {\bf 50}, 1975 (1983).
%
\bibitem{20}
J. Basecq and L. Wolfenstein, Nucl. Phys. B {\bf 224}, 21 (1983).  
%
\bibitem{21}
V. I. Nazaruk, Phys. Lett. B {\bf 229}, 348 (1989).
%
\bibitem{22}
G. N. Fleming, Phys. Lett. B {\bf 125}, 187 (1983).  
%
\bibitem{23}
K. Grotz and H. V. Klapdor, Phys. Rev. C {\bf 30}, 2098 (1984).  
%
\bibitem{24}
P. M. Gopich and I. I. Zaljubovsky, Part. Nuclei {\bf 19}, 785 (1988).
%
\bibitem{25} 
T. Kalogeropouls and G. S. Tzanakos, Phys. Rev. D {\bf 22}, 2585 (1980).
%
\bibitem{26} 
G. S. Mutchlev {\it et al.}, Phys. Rev. D {\bf 38}, 742 (1988). 
%
\bibitem{27}
V. I. Nazaruk, Eur. Phys. J. A {\bf 39}, 249 (2009).  

\end{thebibliography}
\end{document}